\documentclass[preprint]{article}
\usepackage{spconf,amsmath,epsfig,gensymb,breqn,booktabs,multirow}

\AtBeginDocument{
  \setlength\abovedisplayskip{0pt}
  \setlength\belowdisplayskip{0pt}}
  
\title{JOINT SOIL AND ABOVE-GROUND BIOMASS CHARACTERIZATION USING RADARS}
\fourauthors
  {Luke Jacobs}
  {UIUC \\ Electrical and Computer \\	Engineering, Urbana, IL}
  {Mohamad Alipour}
  {UIUC \\ Civil Engineering \\ Urbana, IL}
  {Adam Watts}
  {U.S. Forest Service \\	Research and Development \\ Wenatchee, WA}
  {Elahe Soltanaghai}
  {UIUC \\ Computer Science \\ Urbana, IL}

\begin{document}

\maketitle
\begin{abstract}
Soil moisture sensing through biomass or vegetation canopy has challenged researchers, even those who use SAR sensors with penetration capabilities. This is mainly due to the imposed extra time and phase offsets on Radio Frequency (RF) signals as they travel through the canopy. These offsets depend on the vegetation canopy moisture and height, both of which are typically unknown in agricultural and forest fields. 
In this paper, we leverage the mobility of an unmanned aerial system (UAS) to collect spatially-diverse radar measurements, enabling the joint estimation of soil moisture, above-ground biomass moisture, and biomass height, all without assuming any calibration steps. We leverage the changes in time-of-flight (ToF) and angle-of-arrival (AoA) measurements of reflected radar signals as the UAS flies above a reflector buried under the soil. We demonstrate the effectiveness of our algorithm by simulating its performance under realistic measurement noises as well as conducting lab experiments with different types of above-ground biomass. Our simulation results conclude that our algorithm is capable of estimating volumetric soil moisture to less than 1\% median absolute error (MAE), vegetation height to 11.1cm MAE, and vegetation relative permittivity to 0.32 MAE. Our experimental results demonstrate the effectiveness of the proposed method in practical scenarios for varying biomass moistures and heights.
\end{abstract}
\textit{Copyright 2023 IEEE. Published in the 2023 IEEE International Geoscience and Remote Sensing Symposium (IGARSS 2023), scheduled for 16 - 21 July, 2023 in Pasadena, California, USA. Personal use of this material is permitted. However, permission to reprint/republish this material for advertising or promotional purposes or for creating new collective works for resale or redistribution to servers or lists, or to reuse any copyrighted component of this work in other works, must be obtained from the IEEE. Contact: Manager, Copyrights and Permissions / IEEE Service Center / 445 Hoes Lane / P.O. Box 1331 / Piscataway, NJ 08855-1331, USA. Telephone: + Intl. 908-562-3966.}
\section{INTRODUCTION}
\label{sec:intro}
Soil moisture sensing plays a vital role in numerous agricultural and environmental monitoring applications, from crop monitoring \cite{delong_active_sm_sensing} and irrigation control \cite{cardenas2010precision} in precision agriculture to wildfire risk assessment in forestry environments \cite{chaparro2016predicting,krueger2015soil}. Existing solutions leverage passive and active remote sensing systems to estimate soil permittivity and derive soil moisture. However, the majority of agricultural and forestry sites are covered by canopy layer, rendering remote sensing solutions ineffective. Typical active remote sensing systems \cite{nico_active_sm_sentinel,luo2019uav,liu2020combined,zribi2019analysis,le2002soil} leverage satellite or UAS equipped with radars to associate the received signal strength or the Time of Flight (ToF) of the backscattered signal from the ground to soil moisture levels. However, without calibration aids, these systems cannot separate the effect of surface-layer vegetation from soil moisture estimates \cite{brancato}. This is because the signal must penetrate through the vegetation to reach the soil, the moisture content and height of which will contribute proportionally to the ToF and absorption of radar signals. This makes the dissection of the individual contributions of soil and biomass much more challenging.

Previous works leverage Ultra-Wideband (UWB) radars at C-band combined with either transmission line models \cite{pramudita_teaplant,sinchi2023under} or phase information from an antenna array \cite{comet} to compensate for the vegetation effect. However, these works either make an assumption about the height of vegetation \cite{pramudita_teaplant} or assume a limited vegetation layer height \cite{uav_ir_uwb_drone, comet} that is not applicable in agricultural settings. On the other hand, multi-polarization radars are also used in satellite SAR systems to separate the scattering due to vegetation reflections from the soil surface reflections \cite{brancato,sarker2012forest,mandal2019joint}. However, they are also prone to errors due to atmospheric disturbances and sub-surface scattering effects.

\begin{figure}
    \centering
    \includegraphics[width=0.45\textwidth]
    {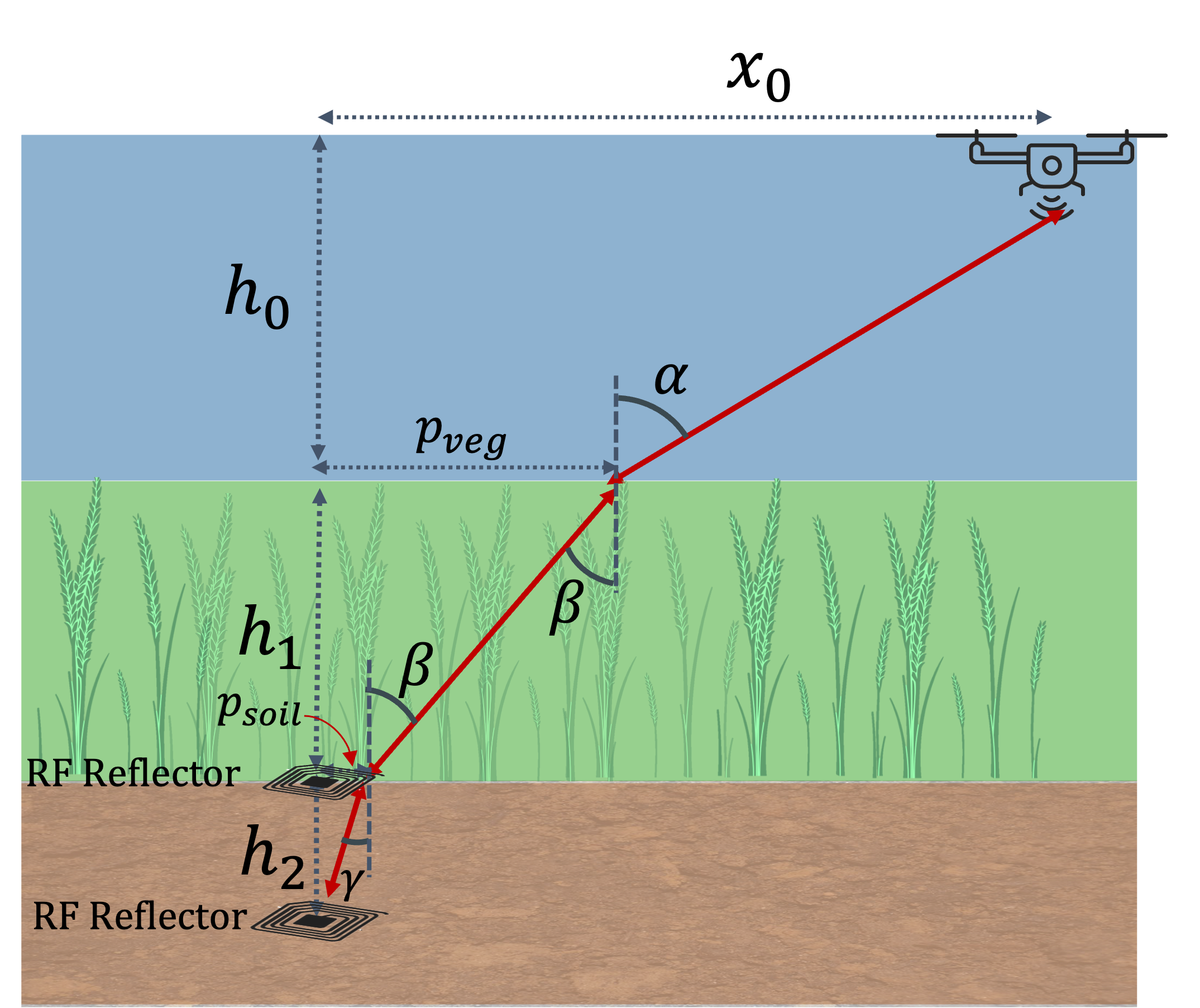}
    \caption{Leveraging radar mobility and the effect of refraction to jointly estimate soil and above-ground biomass characteristics}
    \label{fig:radar_scene}
\end{figure}

Another challenge in estimating soil moisture through canopy originates from the lack of ground references to mark the source of reflections. Existing SAR systems which do not use on-ground infrastructure cannot be certain that the strongest radar reflections originate from the biomass, surface, or ground. This can significantly vary the outcome of remote soil moisture algorithms due to multipath and scattering effects that could potentially mask the reflections from underground and soil surface. To address this challenge, we propose an RFID tag design, based on our previous work \cite{soltanaghaei2021tagfi,soltanaghaei2021millimetro}. The tag acts as a ground reference and marks the reflections penetrated through biomass and soil by modulating the radar reflection. We install two of these RF reflectors, one on the soil surface and one buried in the soil at a known depth, to separate the radar reflections that carry the biomass and soil effects (shown in Figure \ref{fig:radar_scene}). Practically, this installation can be done at the start of a field's growing season, after sowing. 

We propose a radar sensing algorithm that fuses spatially-diverse ToF measurements of the two ground reflectors to jointly infer soil and biomass characteristics without making assumptions about the above-ground biomass properties or requiring calibration steps. Unlike the previous terrestrial soil sensing methods that use radar measurements from only one location \cite{comet,pramudita_teaplant}, we leverage the movement of the radar, either mounted on a drone or agricultural machinery, to collect ToF measurements at multiple locations. We then apply Snell's law of refraction to build a system of equations from independent ToF measurements at different azimuth positions. Next, we employ an iterative non-linear solver to jointly estimate 3 parameters: vegetation height, vegetation permittivity, and soil permittivity. Finally, we can infer soil moisture from soil permittivity using the Topp equation \cite{topp_equation}.

We evaluate our proposed system using simulation with realistic noise parameters as well as lab experimentation using a custom-built step-continuous-wave radar operating at 1.5-2.5GHz and custom-designed RF reflectors. The results demonstrate the effectiveness of this method in accurately estimating soil moisture in the presence of different above-ground biomass and canopy vegetation. We also demonstrate the effectiveness of radar mobility in characterizing the biomass height and permittivity. Next sections elaborate our methodology and obtained results. 

\section{METHODOLOGY}
\label{sec:algorithm_design}

\begin{figure}
    \centering
    \vspace{-2em}
    \includegraphics[width=0.4\textwidth]
    {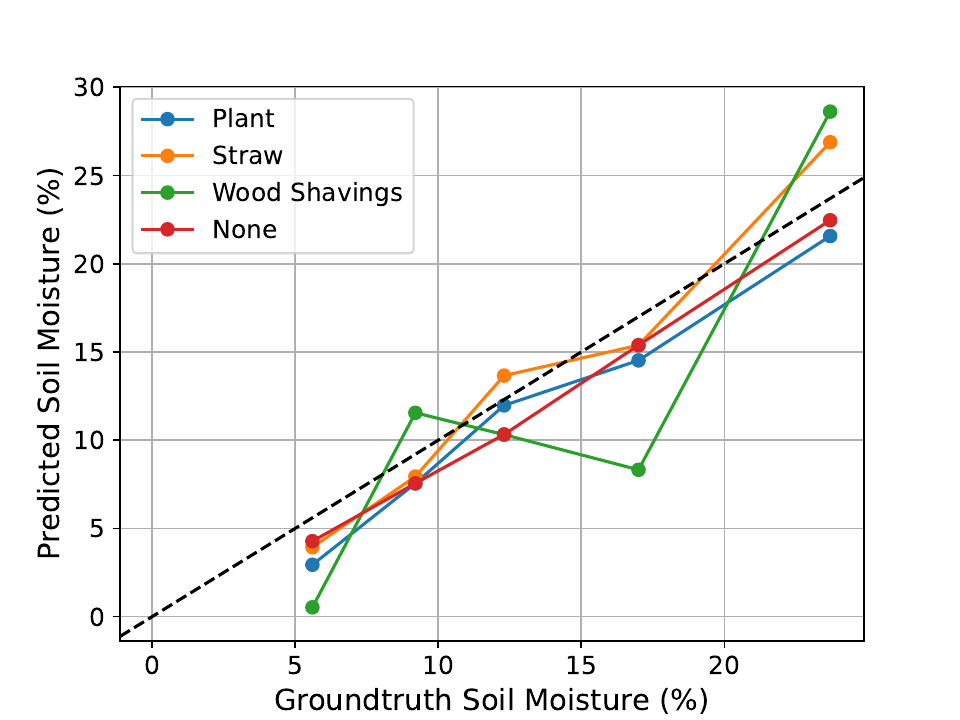}
    \caption{Our experimental evaluations confirm the capability of our joint soil and biomass characterization for different above-ground biomass materials}
    \vspace{-.6em}
    \label{fig:exp_fig_1}
\end{figure}

The proposed system consists of a portable radar, flying or moving above the canopy layer in a forestry or agricultural field, and passive RF reflectors that are sparsely deployed on and under ground as references. As the radar flies above the field of interest, it transmits and receives the RF signals and measures the ToF and angle-of-arrival of the reflected signals. While the proposed method is independent of the radar waveform or the antenna geometry, we perform our experimentation using a stepped frequency continuous-wave (SFCW) radar operating at 1.5 to 2.5GHz. We obtain basic FFT-based signal processing techniques \cite{soltanaghaei2021millimetro} to extract the tag reflections and estimate their ToFs. Without loss of generality, we assume homogeneous layers of air, vegetation, and soil between the ground reflectors and the radar (shown in Figure \ref{fig:exp_fig_1}). It is worth noting that we do not make any assumptions about the above-ground biomass height, $h_1$, or vegetation permittivity, $\epsilon_1$, given that they can vary over time as the growing season progresses \cite{pramudita_teaplant}.

In this model, the soil layer extends from the depth of the buried reflector to the base of the vegetation layer, a height of $h_2$ and permittivity of $\epsilon_2$. While the soil permittivity is one of the unknown variables, the soil depth can be recorded during the tag installation. Finally, it is assumed that the radar is located above the biomass layer at altitude of $h_0+h_1$ from the ground, where $h_0$ defines the air layer height and $h_1$ is the biomass height. While neither of these individual heights are known, we assume that the drone is equipped with navigational sensors capable of measuring the overall altitude.



As the radar signal propagates from the radar TX antenna towards the ground, it will vary in speed $v = c/\sqrt{\epsilon}$ as it passes through new layers of dielectric, where $c$ defines the speed of light and $\epsilon$ defines the relative permittivity of the corresponding layer. At each dielectric interface, some of the signal energy will scatter back towards the radar, producing peaks in the time domain reflectrometry (TDR) output of the radar capture. We place a passive RF reflector at the dielectric interface of biomass and soil to accurately distinguish the signal reflections at this interface from lower layers in the TDR signal. According to Snell's Law, part of the signals at each dielectric interface refracts at an angle $\theta_i$:
\begin{equation}
    n_1 \sin(\theta_1) = n_2 \sin(\theta_2)
    \label{eq:1}
\end{equation}
where $n_i$ defines the relative permitivity of each layer at the interface. 
 We can see that the refracted angle inside the new medium is a function of the incident angle and the permittivities of the adjoined layers. As such, the signal path can be modeled as a series of adjoined line segments of length $l_i$. Once the signal impinges on the reflector, it will scatter in many directions, but according to the law of reciprocity \cite{deepak_dielectric}, the signal that will be received by the radar will be the one that will backtrack along the same path that brought it to the reflector. Therefore, the reflection path is the same as the transmission path. Pulling everything together, the one-way ToF of a radar signal can be defined as: 
\begin{equation}
    \tau = \frac{1}{c} \sum^{N-1}_{i=0}{l_i n_i}
    \label{eq:2}
\end{equation}

\begin{figure*}
    \centering
    \includegraphics[width=0.24\textwidth]{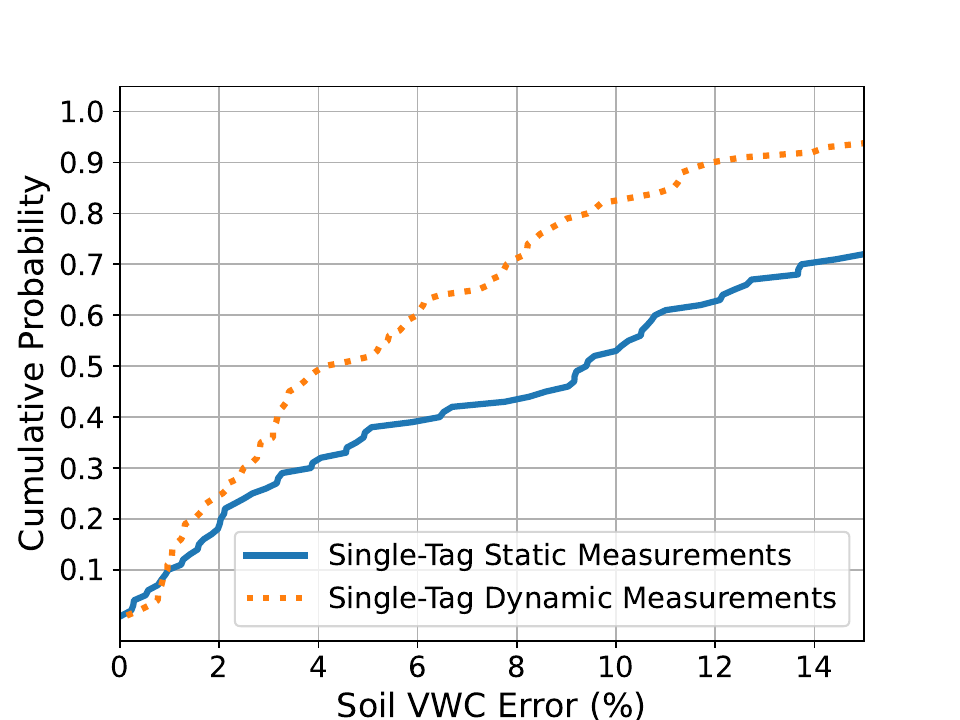}
    \includegraphics[width=0.24\textwidth]{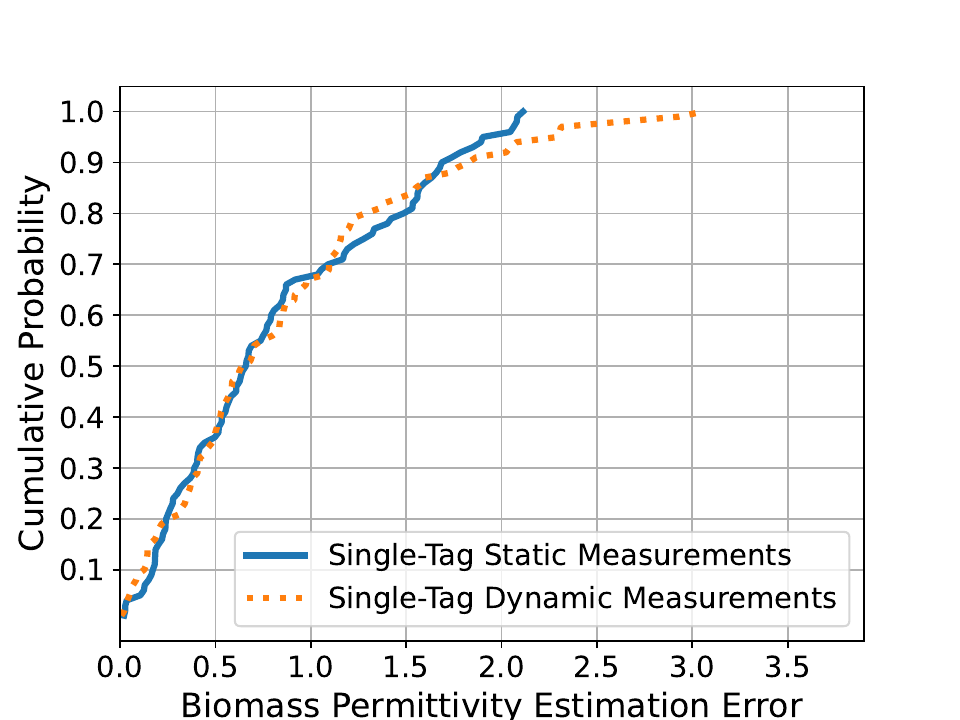}
    \includegraphics[width=0.24\textwidth]{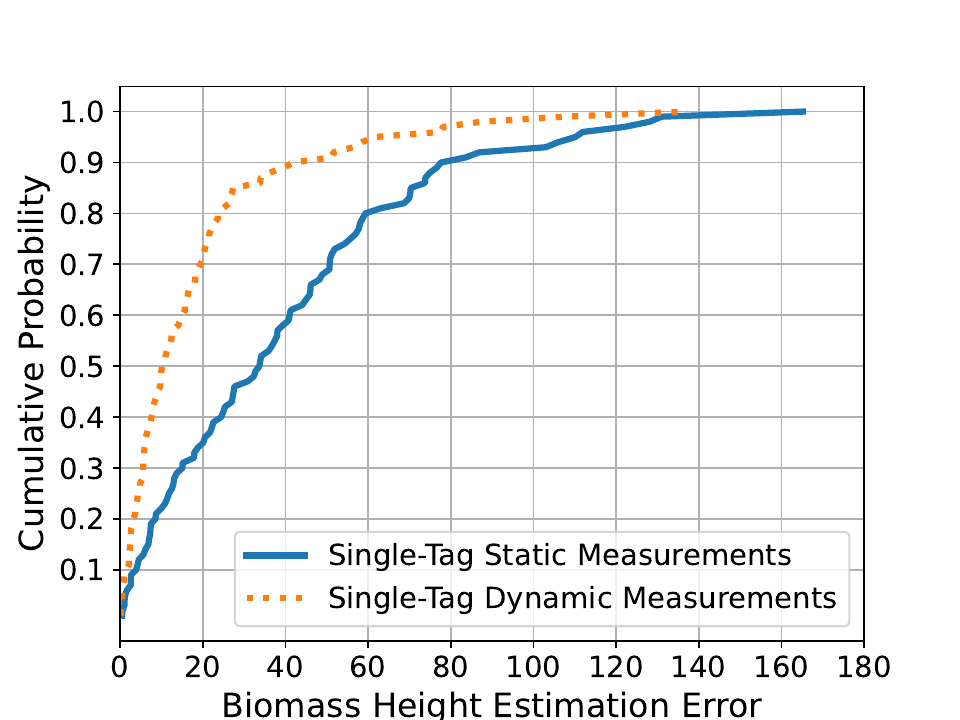}
     \includegraphics[width=0.24\textwidth]
    {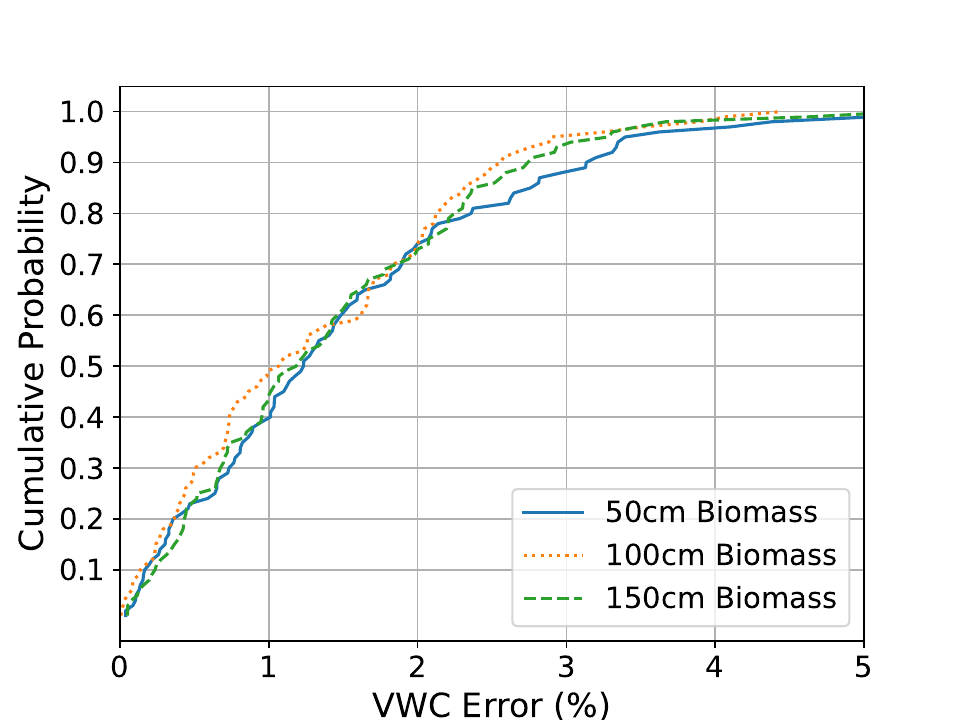}
    \caption{The radar mobility improves the performance of joint soil and biomass characterization (3 left plots), it maintains the high performance regardless of the above-ground biomass height (right plot)}
    \label{fig:cdfSingle}
\end{figure*}

To express this in terms of measurable and unknown quantities, we expand $l_i$ term by defining the two "pierce points," the points at which the radar signal crosses a dielectric interface. Both pierce points, $p_{veg}$ and $p_{soil}$, shown in Fig~\ref{fig:radar_scene}, are defined relative to the normal line originating from the deployed RF reflectors. In order for the radar signal to reach the reflector, the radar signal first enters the vegetation layer at
\begin{align}
    p_{veg} = x_0 - h_0 \tan(\alpha)
    \label{eq:3}
\end{align}
where $\alpha$ is the angle of incidence of the radar transmission or the angle-of-arrival (AoA) of the radar reflection. Note that this angle is different from the angle that would be formed if the signal traveled along the shortest path between the drone and the reflector due to the refraction effects. 
The AoA $\alpha$ can be estimated using an antenna array at the radar or it can be approximated using the known location of the tag and the radar using alternative sensors such as GPS. Our simulation results will elaborate on the impact of $\alpha$ measurement. 
The second pierce point, $p_{soil}$, can also be derived using basic geometry:
\begin{align}
    p_{soil} &= p_{veg} - h_1 \tan(\beta) \nonumber\\
    &= x_0 - h_0 \tan(\alpha) - h_1 \tan(\beta)
    \label{eq:4}
\end{align}
Combining equations \ref{eq:1} to \ref{eq:4}, we can rewrite $p_{soil}$ as
\begin{equation}
\begin{split}
    p_{soil} = x_0 &- h_0 \tan(\alpha) \nonumber\\
    &- h_1 \tan{\left(\arcsin{\left(\frac{n_1 \sin(\alpha)}{n_2}\right)}\right)}
\end{split}
\end{equation}
Note that both the $p_{veg}$ and $p_{soil}$ expressions comprise a mix of measurable quantities ($x_0$, $\alpha$) and unknown quantities ($h_0$, $n_1$, $n_2$). 
Now that we have defined expressions for the pierce point locations, we can determine the segment lengths through each medium by applying the Pythagorean theorem:
\begin{align}
    l_0 &= \sqrt{h_0^2 + (x_0 - p_{veg})^2} \\
    l_1 &= \sqrt{h_1^2 + (p_{veg} - p_{soil})^2} \\
    l_2 &= \sqrt{h_2^2 + p_{soil}^2}
\end{align}
Expanding these terms using equation~\ref{eq:2},
\begin{equation} \label{eq:expanded_tof}
\begin{split}
    \tau = &\frac{1}{c}\sqrt{h_0^2 + (x_0-p_{veg})^2} \\
    &+ \frac{n_1}{c} \sqrt{h_1^2 + (p_{veg}-p_{soil})^2} \\
    &+ \frac{n_2}{c} \sqrt{h_2^2+p_{soil}^2}
\end{split}
\end{equation}
This nonlinear equation contains all of the unknowns of interest ($h_1$, $n_1$, $n_2$) and the measurable quantities ($\tau$, $\alpha$, $x_0$, $h_0+h_1$). For every new measurement location $x_0$ at altitude $h_0+h_1$, the drone collects a new Time of Flight $\tau$ (and, optionally, AoA $\alpha$), thus producing a new nonlinear equation. By collecting these equations into a nonlinear system, we can solve for the three unknowns.

\begin{table*}
  \centering
  \begin{tabular}{lcccr}
    \toprule
    \multirow{2}{*}{Simulation Settings} & \multicolumn{2}{c}{Noiseless, One Reflector} & \multicolumn{2}{c}{Noisy, Two Reflectors} \\
    \cmidrule(lr){2-3} \cmidrule(lr){4-5}
    & Static & Dynamic & Static & Dynamic \\
    \midrule
    Above-ground biomass permittivity median error & 0.43 & 0.34 & 0.34 & 0.32 \\
    Above-ground biomass height median error (cm) & 18.7 & 5.0 & 13.0 & 11.1 \\
    Soil VWC median error & 5.5\% & 2.8\% & 0.008\% & 0.003\% \\
    \bottomrule
  \end{tabular}
  \caption{3-Parameter Estimation Performance in Simulation}
  \label{tab:sim_summary}
\end{table*}

\section{SIMULATION RESULTS}
\label{sec:simulation_results}

We have simulated the performance of our nonlinear solver across a variety of agricultural environments. For each simulation run, we randomly sample vegetation and soil properties from realistic distributions of each variable. We sample the groundtruth soil permittivity $\epsilon_2$ from the uniform distribution between 2 and 20, corresponding to a VWC between near-0\% and 34.5\%. For vegetation simulation, we sample the biomass height $h_1$ between .2 to 2 meters and biomass permittivity $\epsilon_1$ from a uniform distribution between 1.1 and 5, representing a wide range of biomass types from woody biomass to a plant-based canopy layer. To simulate the radar mobility, we consider the drone hovering above the ground reflectors at 0 displacement and moving horizontally, making measurements every .5m up to 2.5m total horizontal displacement. 

First, we ran our simulator using exact ToFs and AoAs that would be measured by a noiseless radar. We compared the performance of the algorithm using static measurements, collected while the UAV hovers over the reflector, to the performance using dynamic measurements, collected while the UAV approaches the reflector. To show the impact of reflectors, we also compare the results of only a single reflector buried under the soil with a two-reflector scenario, one under the soil and one on the soil surface.  Table~\ref{tab:sim_summary} summarizes the errors in estimating soil Volumetric Water Content (VWC), above-ground biomass permittivity and biomass height. The overall performance of our algorithm across all these simulations are shown in Figure \ref{fig:cdfSingle}.

In single reflector scenario, the soil VWC median error using the dynamic measurements is comparable to that of the static measurements, but we can see that the dynamic measurements significantly improves the characterization of vegetation parameters. The main reason is that the optimization problem is under-constrained with a single static measurement when modeling more than one dielectric layer, which affects the performance in estimating the parameters of the top layer (in this case, the biomass layer). Figure \ref{fig:cdfSingle} also shows the performance of the proposed method in dissecting the soil layer from biomass for different biomass heights. As we can see, the use of the two reflectors allows us to accurately estimate the soil volume water content regardless of the density or height of the biomass layer above the soil.

Second, we ran our simulator using perturbed ToF's that would be similar to the measurements collected by a commodity radar. The measurement noises necessitates the use of the second tag at the surface layer to differentiate the reflections from soil surface and under the soil. Comparing the static and dynamic measurement results in Table \ref{tab:sim_summary} for the noisy scenario, we can see that the dynamic measurement method improves the estimation of vegetation properties over the static measurement method. 
 


\section{EXPERIMENTAL RESULTS and Conclusion}
\label{sec:majhead}
We performed lab-based experiments to show the real-world validity of our solver using a Keysight Streamline Vector Network Analyzer (VNA) \cite{keysight_p5008b_nodate} with step-CW waveform spanning 1.5 to 2.5GHz and a Vivaldi antenna \cite{wa5vjb_wa5vjb_nodate} attached to port 1 of the VNA. We then capture the complex S11 values (single-antenna frequency-domain channel estimates) and perform the Inverse Fast Fourier Transform to extract the TDR signal. We use a match filtering algorithm similar to \cite{soltanaghaei2021millimetro} to detect and separate the reflections from the modulating RF reflectors. 

First, we conducted a set of experiments to demonstrate the robustness of soil moisture estimation in the presence of different above-ground biomass materials. In these experiments, the radar is at a fixed position in front of the reflector at a distance of 1.5m. In front of the reflector, there is a soil layer with a fixed depth of 13cm. We then place different biomass materials of different height in front of soil such as two corn plants, a box of wood shavings with an average particle size of 2.5mm, and a box of straw with 6.4mm particle size. The results are summarized in Fig~\ref{fig:exp_fig_1} and we can see the effectiveness of our proposed method in accurately estimating soil moisture in the presence of above-ground biomass. We can see a small performance degradation in wood shaving scenario, which is mainly due to high density of this biomass material and the resulting reduction in signal to noise ratio.


\begin{figure}
    \centering
    \vspace{-1em}
    \includegraphics[width=0.31\linewidth]
    {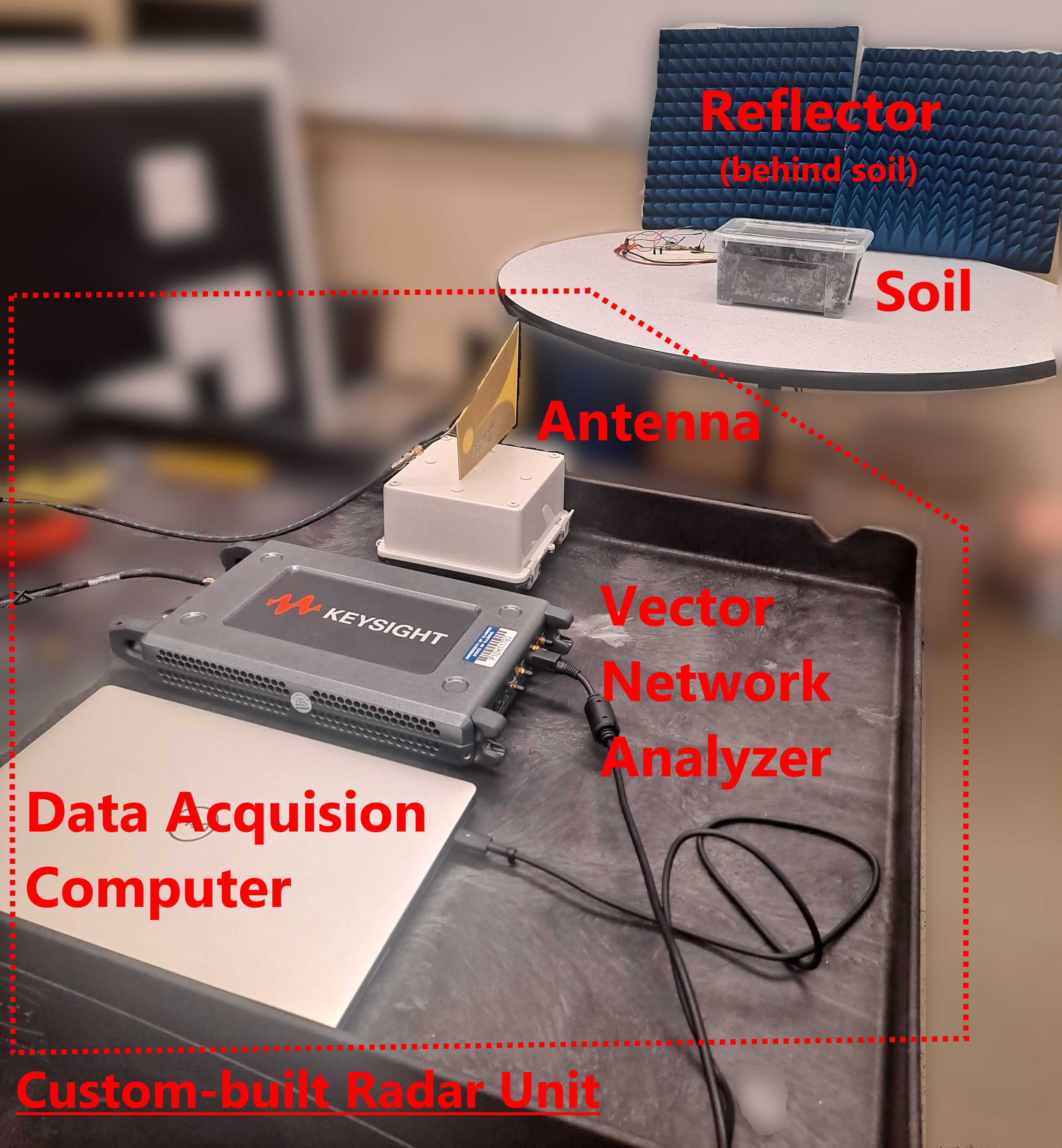}
    \includegraphics[width=0.68\linewidth]
    {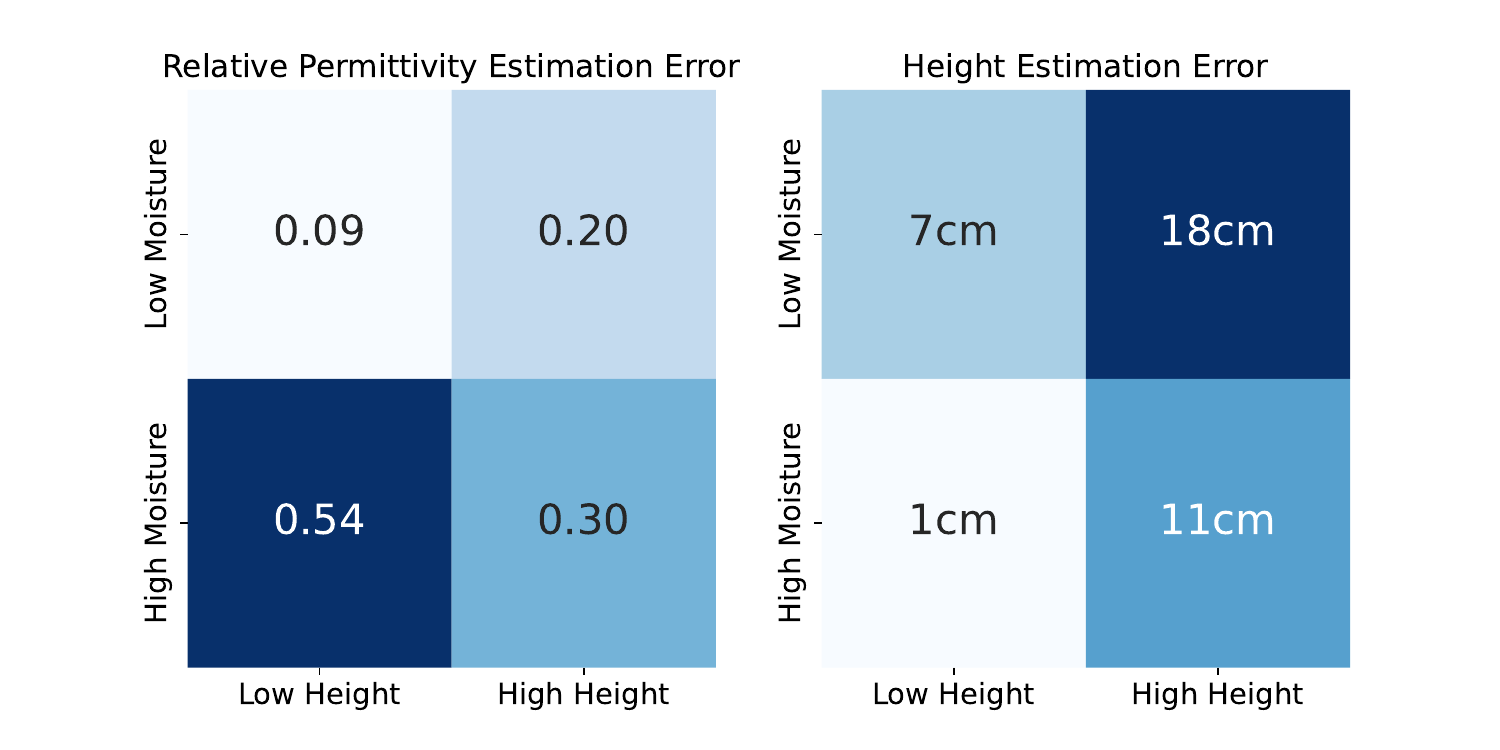}
    \caption{Radar setup for experimental evaluation (left) and the accuracy of above-ground biomass properties estimations in low and high biomass moisture and heights }
    \label{fig:exp_fig_2}
\end{figure}
Next, we show the effectiveness of radar mobility in jointly estimating biomass height and permittivity above the soil. We run 4 different scenarios which represent two biomass moisture levels (measured relative permittivity of 1.51 and 1.89 using the Teros-12 sensor) and two biomass height levels (35cm and 51cm). We fixed the biomass type to wood shavings in all these experiments. To emulate radar movements, we collected ToF measurements at four different horizontal displacements between 0m to 1.0m. The results of these experiments are summarized in Fig~\ref{fig:exp_fig_2}. We can see that the dynamic measurement method can characterize biomass permittivity and height to levels of precision useful to farmers and wildland managers. As part of our future work, we plan to extend these experiments to in-field testing by mounting the radar on a drone.

\bibliographystyle{IEEEbib}
\bibliography{refs}

\begin{thebibliography}{10}

\bibitem{delong_active_sm_sensing}
Kyle DeLong, Brian Hornbuckle, Jun Wang, Michael Cosh, and Daryl Herzmann,
\newblock ``Validating soil moisture with farmers in mind: A new approach for remote sensing and modeling in the us corn belt,''
\newblock in {\em IGARSS 2023 - 2023 IEEE International Geoscience and Remote Sensing Symposium}, 2023, pp. 2669--2672.

\bibitem{cardenas2010precision}
Bernard Cardenas-Lailhacar and Michael~D Dukes,
\newblock ``Precision of soil moisture sensor irrigation controllers under field conditions,''
\newblock {\em Agricultural Water Management}, vol. 97, no. 5, pp. 666--672, 2010.

\bibitem{chaparro2016predicting}
David Chaparro, Merce Vall-Llossera, Maria Piles, Adriano Camps, Christoph R{\"u}diger, and Ramon Riera-Tatch{\'e},
\newblock ``Predicting the extent of wildfires using remotely sensed soil moisture and temperature trends,''
\newblock {\em IEEE journal of selected topics in applied earth observations and remote sensing}, vol. 9, no. 6, pp. 2818--2829, 2016.

\bibitem{krueger2015soil}
Erik~S Krueger, Tyson~E Ochsner, David~M Engle, JD~Carlson, Dirac Twidwell, and Samuel~D Fuhlendorf,
\newblock ``Soil moisture affects growing-season wildfire size in the southern great plains,''
\newblock {\em Soil Science Society of America Journal}, vol. 79, no. 6, pp. 1567--1576, 2015.

\bibitem{nico_active_sm_sentinel}
Giovanni Nico, Olimpia Masci, Nuno~Cirne Mira, João Catalão, and Pedro Mateus,
\newblock ``Estimating soil moisture by sentinel-1, sentinel-2 and prisma data: Assessment of results and comparison with in-situ measurements,''
\newblock in {\em IGARSS 2023 - 2023 IEEE International Geoscience and Remote Sensing Symposium}, 2023, pp. 2645--2648.

\bibitem{luo2019uav}
Wei Luo, Xianli Xu, Wen Liu, Meixian Liu, Zhenwei Li, Tao Peng, Chaohao Xu, Yaohua Zhang, and Rongfei Zhang,
\newblock ``Uav based soil moisture remote sensing in a karst mountainous catchment,''
\newblock {\em Catena}, vol. 174, pp. 478--489, 2019.

\bibitem{liu2020combined}
Ying Liu, Jiaxin Qian, and Hui Yue,
\newblock ``Combined sentinel-1a with sentinel-2a to estimate soil moisture in farmland,''
\newblock {\em IEEE Journal of Selected Topics in Applied Earth Observations and Remote Sensing}, vol. 14, pp. 1292--1310, 2020.

\bibitem{zribi2019analysis}
Mehrez Zribi, Sekhar Muddu, Safa Bousbih, Ahmad Al~Bitar, Sat~Kumar Tomer, Nicolas Baghdadi, and Soumya Bandyopadhyay,
\newblock ``Analysis of l-band sar data for soil moisture estimations over agricultural areas in the tropics,''
\newblock {\em Remote Sensing}, vol. 11, no. 9, pp. 1122, 2019.

\bibitem{le2002soil}
Sylvie Le~H{\'e}garat-Mascle, Mehrez Zribi, F~Alem, A~Weisse, and C{\'e}cile Loumagne,
\newblock ``Soil moisture estimation from ers/sar data: Toward an operational methodology,''
\newblock {\em IEEE Transactions on Geoscience and Remote Sensing}, vol. 40, no. 12, pp. 2647--2658, 2002.

\bibitem{brancato}
Virginia Brancato and Irena Hajnsek,
\newblock ``Separating the influence of vegetation changes in polarimetric differential sar interferometry,''
\newblock {\em IEEE Transactions on Geoscience and Remote Sensing}, vol. 56, no. 12, pp. 6871--6883, 2018.

\bibitem{pramudita_teaplant}
Aloysius~A. Pramudita, Yuyu Wahyu, Syamsul Rizal, Murman~D. Prasetio, Agung~N. Jati, Restu Wulansari, and Harfan~H. Ryanu,
\newblock ``Soil water content estimation with the presence of vegetation using ultra wideband radar-drone,''
\newblock {\em IEEE Access}, vol. 10, pp. 85213--85227, 2022.

\bibitem{sinchi2023under}
Kurt~Soncco Sinchi, Diego Calderon, Ishfaq Aziz, Adam Watts, Elahe Soltanaghai, and Mohamad Alipour,
\newblock ``Under-canopy biomass sensing using uas-mounted radar: a numerical feasibility analysis,''
\newblock in {\em IGARSS 2023-2023 IEEE International Geoscience and Remote Sensing Symposium}. IEEE, 2023, pp. 3292--3295.

\bibitem{comet}
Usman~Mahmood Khan and Muhammad Shahzad,
\newblock ``Estimating soil moisture using rf signals,''
\newblock in {\em Proceedings of the 28th Annual International Conference on Mobile Computing And Networking}, New York, NY, USA, 2022, MobiCom '22, p. 242–254, Association for Computing Machinery.

\bibitem{uav_ir_uwb_drone}
Rong Ding, Haiming Jin, Dong Xiang, Xiaocheng Wang, Yongkui Zhang, Dingman Shen, Lu~Su, Wentian Hao, Mingyuan Tao, Xinbing Wang, and Chenghu Zhou,
\newblock ``Soil moisture sensing with uav-mounted ir-uwb radar and deep learning,''
\newblock {\em Proc. ACM Interact. Mob. Wearable Ubiquitous Technol.}, vol. 7, no. 1, mar 2023.

\bibitem{sarker2012forest}
Md~Latifur~Rahman Sarker, Janet Nichol, Huseyin~Baki Iz, Baharin~Bin Ahmad, and Alias~Abdul Rahman,
\newblock ``Forest biomass estimation using texture measurements of high-resolution dual-polarization c-band sar data,''
\newblock {\em IEEE Transactions on Geoscience and Remote Sensing}, vol. 51, no. 6, pp. 3371--3384, 2012.

\bibitem{mandal2019joint}
Dipankar Mandal, Vineet Kumar, Heather McNairn, Avik Bhattacharya, and YS~Rao,
\newblock ``Joint estimation of plant area index (pai) and wet biomass in wheat and soybean from c-band polarimetric sar data,''
\newblock {\em International Journal of Applied Earth Observation and Geoinformation}, vol. 79, pp. 24--34, 2019.

\bibitem{soltanaghaei2021tagfi}
Elahe Soltanaghaei, Adwait Dongare, Akarsh Prabhakara, Swarun Kumar, Anthony Rowe, and Kamin Whitehouse,
\newblock ``Tagfi: Locating ultra-low power wifi tags using unmodified wifi infrastructure,''
\newblock {\em Proceedings of the ACM on Interactive, Mobile, Wearable and Ubiquitous Technologies}, vol. 5, no. 1, pp. 1--29, 2021.

\bibitem{soltanaghaei2021millimetro}
Elahe Soltanaghaei, Akarsh Prabhakara, Artur Balanuta, Matthew Anderson, Jan~M Rabaey, Swarun Kumar, and Anthony Rowe,
\newblock ``Millimetro: mmwave retro-reflective tags for accurate, long range localization,''
\newblock in {\em Proceedings of the 27th Annual International Conference on Mobile Computing and Networking}, 2021, pp. 69--82.

\bibitem{topp_equation}
G.~C. Topp, J.~L. Davis, and A.~P. Annan,
\newblock ``Electromagnetic determination of soil water content: Measurements in coaxial transmission lines,''
\newblock {\em Water Resources Research}, vol. 16, no. 3, pp. 574--582, 1980.

\bibitem{deepak_dielectric}
Deepak Vasisht, Guo Zhang, Omid Abari, Hsiao-Ming Lu, Jacob Flanz, and Dina Katabi,
\newblock ``In-body backscatter communication and localization,''
\newblock in {\em Proceedings of the 2018 Conference of the ACM Special Interest Group on Data Communication}, New York, NY, USA, 2018, SIGCOMM '18, p. 132–146, Association for Computing Machinery.

\bibitem{keysight_p5008b_nodate}
Keysight,
\newblock ``P5008b streamline vector network analyzer, 100 {kHz} to 53 {GHz}, 2-port,'' https://www.keysight.com/us/en/product/P5008B/streamline-vector-network-analyzer-100-khz-to-53-ghz-2-port.html.

\bibitem{wa5vjb_wa5vjb_nodate}
WA5VJB,
\newblock ``{WA}5{VJB} antennas,'' https://www.wa5vjb.com/products5.html.

\end{thebibliography}

\end{document}